\documentclass{ckm}                 % twocolumn proceedings style
\usepackage{txfonts}            % if you have it, to get Times family
\confname{Workshop on the CKM Unitarity Triangle, IPPP Durham, April 2003}

\title{Estimate of SU(3) flavour symmetry breaking in 
$\bar B_s\to K^+ K^-$ vs. $\bar B_d\to\pi^+\pi^-$ decay}

\author{M Beneke}
\address{Institut f\"ur Theoretische Physik E, RWTH Aachen, 
Sommerfeldstr. 28, D - 52074 Aachen}
\begin{document}

\begin{abstract}
\vskip-4.5cm
{\normalsize PITHA 03/04, hep-ph/0308040, 27 June 2003}
\vskip3.8cm
I estimate the SU(3) flavour symmetry breaking in the 
ratio of penguin-to-tree ratios of the decays 
$\bar B_s\to K^+ K^-$ and $\bar B_d\to\pi^+\pi^-$, given 
an assumption on the flavour symmetry breaking of the 
hadronic input parameters. The decay amplitudes are calculated 
in QCD factorization. Implications for the determination of 
$\gamma$ are discussed. 
\end{abstract}

\maketitle

\section{Introduction}

In this note I perform a study 
of SU(3) breaking for the decays $\bar B_d\to\pi^+\pi^-$ and 
$\bar B_s\to K^+ K^-$. This system is interesting, because it allows for 
a determination of the angle $\gamma$ from mixing-induced and direct CP 
asymmetries \cite{Dunietz:1993rm,Fleischer:1999pa} provided the SU(3) 
symmetry breaking corrections to a certain 
double ratio of amplitudes are known. The analysis of SU(3) breaking 
is done in the theoretical framework of QCD factorization
\cite{BBNS1,BBNS3,PV}, which expresses the hadronic decay amplitudes in 
terms of fundamental constants, decay constants, form factors etc. 
in the heavy quark limit. The following is a preliminary version of 
work in progress in collaboration with M. Neubert. 

We write the decay amplitudes as
\begin{eqnarray}
   && A(\bar B_d\to\pi^+\pi^-) 
   = \frac{i \,G_F}{\sqrt2} \left( \lambda_u^{(d)}\,T_\pi
    + \lambda_c^{(d)}\,P_\pi \right) \,, 
\nonumber\\
   && A(\bar B_s\to K^+ K^-) 
   = \frac{i \,G_F}{\sqrt2} \left( \lambda_u^{(s)}\,T_K
    + \lambda_c^{(s)}\,P_K \right)\,
\end{eqnarray}
where $\lambda^{(D)}_p=V_{pb} V^*_{pD}$. 
The ``tree'' and ``penguin'' amplitudes are defined as the coefficients 
of the two terms with different CKM matrix elements. They also contain 
subleading penguin, electroweak penguin and weak annihilation amplitudes. 
In the SU(3) symmetry limit 
$T_\pi=T_K$ and $P_\pi=P_K$. We will be interested in the deviations of 
the ratios
\begin{equation}
   r_T\equiv \frac{T_K}{T_\pi} \,, \qquad 
   r_{PT}\equiv \frac{P_K/T_K}{P_\pi/T_\pi}
\end{equation}
from unity. 

The ratio of tree amplitudes is the product of a factorizable and a 
non-factorizable term, $r_T\equiv r^{\rm f}\,r_T^{\rm nf}$. The
factorizable term is given by
\begin{eqnarray}
   r^{\rm f}
   &=& \frac{(1-m_K^2/m_{B_s}^2)\,(1-4 m_K^2/m_{B_s}^2)^{1/2}}
            {(1-m_\pi^2/m_B^2)\,(1-4 m_\pi^2/m_B^2)^{1/2}}\,
    \frac{f_K}{f_\pi}\,\frac{F_0^{B_s\to K}(m_K^2)}
                            {F_0^{B\to\pi}(m_\pi^2)} \nonumber\\
   &=& 1.19\times\frac{F_0^{B_s\to K}(m_K^2)}{F_0^{B\to\pi}(m_\pi^2)} 
   \approx 1.27\,,
\end{eqnarray}
where we have included the light meson masses and phase space
effects. The $B\to\pi$ form factor is taken from \cite{ball}, but 
the assumed value for the $B_s\to K$ form factor is only 
an ``educated guess''. In the context of QCD sum rules this form 
factor is expected to depend sensitively on the poorly known first 
Gegenbauer moment of the kaon light-cone distribution amplitude, 
since the form factor is dominated by the soft 
spectator-quark overlap term.

\section{Estimate of SU(3) breaking}

The factorizable SU(3) breaking correction cancels in the double ratio 
$r_{PT}$ \cite{Fleischer:1999pa}. This ratio and $r_T^{\rm nf}$ 
deviate from unity due to non-factorizable effects that can be 
computed in QCD factorization assuming that $m_b\gg \Lambda_{\rm
  QCD}$. The computation is a straightforward extension of the 
results of \cite{BBNS3}. I refer to this paper and \cite{PV} 
for all details concerning the method and the values of the hadronic 
parameters. The parameters relevant to SU(3) breaking are:
\begin{itemize}
\item[--] 
$r_\chi^K$ vs.\ $r_\chi^\pi$ in the normalization of scalar penguin 
terms;
\item[--] 
the ratio $f_{B_s} f_K/(m_{B_s}^2\,F_0^{B_s\to K})$ vs.\ 
$f_{B} f_\pi/(m_{B}^2\,F_0^{B\to \pi})$ in the normalization of the 
hard-scattering and weak annihilation terms;
\item[--] 
the Gegenbauer moments $\alpha_1^K$, $\alpha_2^K$ vs.\ $\alpha_1^\pi=0$, 
$\alpha_2^\pi$, and the first inverse moments of the $B$-meson 
distribution amplitudes, $m_B/\lambda_B$ vs.\ $m_{B_s}/\lambda_{B_s}$;
\item[--] 
the parameters for soft power corrections from hard scattering and weak 
annihilation, $X_A$ and $X_H$.
\end{itemize}
Since these parameters are often not well-known, we will make assumptions 
on the SU(3) breaking and exhibit the effect of these assumptions on the 
ratios $r_T^{\rm nf}$ and $r_{PT}$. This should be distinguished from 
more ambitious approaches as discussed in \cite{khod}, where the 
SU(3) breaking in the input parameters is also computed. 

For a given pair of related parameters denoted $(x_\pi,x_K)$, for 
instance $(x_\pi,x_K)=(\lambda_B,\lambda_{B_s})$, we set all other parameters 
and $x_\pi$ to their standard values. We then write 
$x_K=x_\pi\,(1+\delta)$ and vary $\delta$, which controls the amount of
SU(3) breaking, between $-0.3$ and 0.3. For the complex parameters $X_H$ 
and $X_A$ we vary the magnitude and phase simultaneously and 
independently by this amount. We make two exceptions to this treatment. 
For the light meson Gegenbauer moments, which are already small 
corrections to the SU(3) symmetric asymptotic distribution amplitudes, we 
take $\alpha_1^{\bar K}=0.2\pm 0.2$, $\alpha_2^{\bar K}=0.1\pm 0.3$, and 
$\alpha_2^\pi=0.1$. Second, we use
\begin{equation}
   \frac{r_\chi^K}{r_\chi^\pi}
   = \frac{2 m_K^2}{m_\pi^2}\,\frac{1}{m_s/m_q+1} = 0.99\pm 0.06
\end{equation}
with $m_s/m_q=24.2\pm 1.5$ taken from \cite{Leutwyler:1996qg}. 

The resulting variations of the magnitudes and phases of $r_T^{\rm nf}$ 
and $r_{PT}$ about their default values 
\begin{equation}\label{rPTdef}
   r_T^{\rm nf} = 0.99\,e^{0.2^\circ i} \,, \qquad
   r_{PT} = 1.02\,e^{-1.1^\circ i}
\end{equation}
are shown in the upper part of Table~\ref{tab:su3}. Adding up the various 
uncertainties we obtain non-factorizable SU(3) breaking effects of order 
$\pm 5\%$ for the ratio of tree amplitudes and of order $\pm 10\%$ for 
the double ratio of penguin and tree amplitudes. Since the strong phases 
of the ratios are very small, the SU(3) breaking effect on the phases is 
also small in absolute magnitude. 

These estimates have to be regarded with caution, since they rely on the 
default estimate of weak annihilation ($\varrho_A=0$), in which the 
annihilation amplitude does not have a strong phase. To study the impact 
of weak annihilation in more detail we pick the four values $X_A^i$ 
corresponding to 
$|\varrho_A|=0.75$ with phase 0, 90, 180, 270 degrees for the pion mode 
and allow the corresponding parameters to vary by $\pm 30\%$ for the kaon 
mode. All the values of the annihilation parameter chosen by this 
procedure lie (almost) 
within our standard error assumption $|\varrho_A|<1$. The 
lower part of Table~\ref{tab:su3} demonstrates that the SU(3) breaking 
effect on the ratio of penguin amplitudes can be large, up to $30\%$ in 
magnitude and $\pm 15^\circ$ on the phase of $r_{PT}$. We conclude that 
unless a better understanding of SU(3) breaking effects in the weak 
annihilation amplitude is found the assumption $r_{PT}=1$ should not be 
made. 

\begin{table}
\caption{\label{tab:su3}
SU(3) breaking effect on the magnitude and phase of $r_T^{\rm nf}$ and 
$r_{PT}$ for variations of a given input parameter as described in the 
text.}
\vspace{0.1cm}
\begin{center}
\begin{tabular}{|c|c|c|c|c|}
\hline\hline
&&&& \\[-0.3cm]
Parameter & $\delta\,|r_T^{\rm nf}|$
 & $\delta\,\mbox{arg}(r_T^{\rm nf})$
 & $\delta\,|r_{PT} |$ & $\delta \,\mbox{arg}(r_{PT})$ \\[0.1cm]
\hline
&&&& \\[-0.3cm]
$r_\chi$ & $\pm 0.01$ & $<0.1^\circ$ & $\pm 0.05$ & $<0.1^\circ$ \\[0.1cm]
$F_0^{B\to M}$ & ${}_{-0.03}^{+0.02}$ & $<0.1^\circ$
 & ${}_{-0.06}^{+0.05}$ & $<0.5^\circ$ \\[0.1cm]
$\lambda_B$ & ${}_{-0.04}^{+0.02}$ & $<0.3^\circ$ & ${}_{-0.02}^{+0.03}$
 & $<0.1^\circ$ \\[0.1cm]
$\alpha_1^M$ & $\pm 0.01$ & $<0.1^\circ$ & $\pm 0.02$ & $\pm 1^\circ$ \\[0.1cm]
$\alpha_2^M$ & ${}_{-0.04}^{+0.03}$ & $\pm 0.2^\circ$
 & ${}_{-0.00}^{+0.01}$ & $ <0.1^\circ$ \\[0.1cm]
$X_H$ & $\pm 0.01$ & $<0.1^\circ$ & $\pm 0.01$ & $ <0.1^\circ$ \\[0.1cm]
$X_A\,(\varrho_A^\pi=0)$ & ${}_{-0.01}^{+0.00}$ & $<0.1^\circ$
 & ${}^{+0.10}_{-0.07}$ & $\pm 0.5^\circ$ \\[0.1cm]
\hline
&&&& \\[-0.3cm]
Sum & ${}_{-0.06}^{+0.04}$ & $\pm 0.2^\circ$ & ${}_{-0.10}^{+0.13}$
 & $\pm 1.2^\circ$ \\[0.1cm]
\hline
&&&& \\[-0.3cm]
$X_A^1$ & $\pm 0.00$ & $<0.1^\circ$
 & ${}^{+0.29}_{-0.13}$ & $\pm 1^\circ$ \\[0.1cm]
$X_A^2$ & $\pm 0.01$ & $<0.5^\circ$
 & ${}^{+0.19}_{-0.10}$ & ${}^{+13^\circ}_{-5^\circ}$ \\[0.1cm]
$X_A^3$ & $\pm 0.00$ & $<0.1^\circ$
 & ${}_{-0.03}^{+0.00}$ & $<0.5^\circ$ \\[0.1cm]
$X_A^4$ & $\pm 0.01$ & $<0.5^\circ$
 & ${}^{+0.18}_{-0.09}$ & ${}^{+5^\circ}_{-14^\circ}$ \\[0.1cm]
\hline\hline
\end{tabular}
\end{center}
\end{table}

\section{Determination of $\gamma$}

To estimate the theoretical uncertainty in the determination of the angle
$\gamma$ due to the SU(3) breaking effects, we assume that the mixing 
phase $\phi_{B_d}=2\beta$ is $47^\circ$, corresponding to 
$\sin(2\beta)=0.731$, and define the time-dependent CP asymmetry through 
\begin{eqnarray}
   A_{\rm CP}(t)
   &=& \frac{\mbox{Br}(B(t)\to f)-\mbox{Br}(\bar B(t)\to\bar f)}
          {\mbox{Br}(B(t)\to f)+\mbox{Br}(\bar B(t)\to\bar f)} 
\nonumber\\[0.2cm]
   &=& - S_f \sin(\Delta m_B\,t) + C_f \cos(\Delta m_B\,t) \,.
\end{eqnarray}
We assume that $S_{\pi\pi}=-0.6$ and $C_{\pi\pi}=-0.1$ have been 
measured. We can then determine 
\begin{equation}
   d e^{i\theta}\equiv
   \left|\frac{\lambda_c^{(d)}}{\lambda_u^{(d)}}\right|\,
   \frac{[-P_\pi]}{T_\pi}
\end{equation}
as a function of $\gamma$,\footnote{The definition of $d e^{i\theta}$ 
differs from \cite{Fleischer:1999pa} by a sign, so that $\theta$ is near 
zero in QCD factorization.} 
and predict $C_{KK}$, the direct CP asymmetry in $B_s\to K^+ K^-$ decay, 
for a given assumed value of the SU(3) breaking ratio $r_{PT}$. A 
measurement of $C_{KK}$ then results in a determination of $\gamma$ with 
a theoretical error due to the uncertainty of $r_{PT}$. 

\begin{figure}
\hspace*{-1.3cm}
\hbox to\hsize{\hss
\includegraphics[width=10cm]{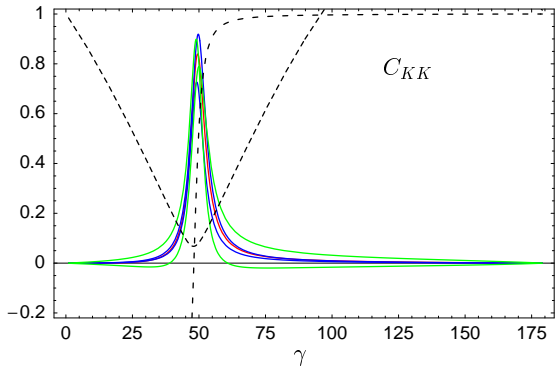}
\hss}
\caption{
The direct CP asymmetry in $B_s\to K^+ K^-$ as a function of $\gamma$ for 
five values of $r_{PT}$: $1.02\,e^{-1^\circ i}$ (dark/red), 
$1.3$, $0.85$, (dark/blue) $e^{15^\circ i}$ and 
$e^{-15^\circ i}$ (light/green). The dashed lines show $d$ and 
$\cos\theta$. $\cos\theta$ changes rapidly around $\gamma=50^\circ$.}
\label{fig:su3}
\end{figure}

We dismiss solutions with $d>1$ as unphysical, since a large penguin 
amplitude is or will be excluded by branching fraction measurements. We 
generically find two solutions for $d e^{i\theta}$. The second solution 
satisfies $d<1$ only in a small range around $\gamma=170^\circ$ and will 
not be discussed further. The other solution already constrains 
$5^\circ<\gamma<95^\circ$ from $d<1$ alone. The question is how a 
measurement of $C_{KK}$ narrows this range further. In 
Figure~\ref{fig:su3} we show $C_{KK}$ as a function of $\gamma$ for five 
values of $r_{PT}$: the central value according to (\ref{rPTdef}), and 
for the most pessimistic error assumptions corresponding to $1.3$, 
$0.85$, $e^{15^\circ i}$ and $e^{-15^\circ i}$. The figure also shows $d$ 
and $\cos\theta$. 

The CP asymmetry exhibits a resonance-like behaviour after $d$ has gone 
through its minimal value, since the denominator of the expression for 
$C_{KK}$ goes through a minimum when $d\cos\theta$ becomes small and 
positive. We find that the largest error is introduced by the uncertainty 
in the phase of $r_{PT}$, which in our analysis is entirely due to weak 
annihilation. For instance, if $C_{KK}=0.06$ is found, the strategy would 
determine two values of $\gamma$ in the absence of theoretical (and 
experimental) errors. Including the SU(3) breaking error, we obtain
$\gamma=(40^{+3}_{-6})^\circ$ and $\gamma=(65^{+13}_{-9})^\circ$. Since 
$\cos\theta$ is expected to be positive, the second range is 
theoretically favored. In general, the theoretical error can become 
larger or smaller depending on $C_{KK}$ as seen from the figure. It will 
also depend on what the mixing-induced and direct CP asymmetries in 
$B_d\to\pi^+\pi^-$ will eventually turn out to be, but our result should 
be generic.  

We therefore conclude that the strategy to determine $\gamma$ from the 
$B_d\to\pi^+\pi^-$ to $B_d\to K^+ K^-$ system {\em may\/} suffer from 
considerable theoretical uncertainties, unless additional information 
is available. This can come from two sources: (1) Excluding the 
possibility of a weak annihilation amplitude with a large strong 
rescattering phase would already eliminate the largest uncertainty 
that we could identify in the context of QCD factorization. 
This could be achieved experimentally by excluding a large direct 
CP asymmetry in $B\to\pi^- K^+$ and neglecting SU(3) breaking 
effects in the estimate of the weak annihilation phase. (2) 
If the $B_s \bar B_s$ mixing phase is known, the mixing-induced CP 
asymmetry in $B_s\to K^+ K^-$ decay provides a fourth observable, 
which could be used to determine (or eliminate) $\theta$, leaving 
only the SU(3) breaking error on $d$. To make this work in practice, 
a significant experimental effort is required.

\end{document}